# On-Demand Mobility Services for Infrastructure and Community Resilience: A Review toward Synergistic Disaster Response Systems

**Jiangbo Yu, Matthew Korp**

Email: jiangbo.yu@mcgill.ca
Phone: (514) 702-5675

Macdonald Engineering Building
817 Sherbrooke Street West, Room 475C
Department of Civil Engineering
McGill University
Montreal, QC, Canada H3A 0C3

## Abstract

Mobility-on-demand (MOD) services have the potential to significantly improve the adaptiveness and recovery of urban systems, in the wake of disruptive events. But there lacks a comprehensive review on using MOD services for such purposes in addition to serving regular travel demand. This paper presents a review that suggests a noticeable increase within recent years on this topic across four main areas – resilient MOD services, novel usage of MOD services for improving infrastructure and community resilience, empirical impact evaluation, and enabling and augmenting technologies. The review shows that MOD services have been utilized to support anomaly detection, essential supply delivery, evacuation and rescue, on-site medical care, power grid stabilization, transit service substitution during downtime, and infrastructure and equipment repair. Such a versatility suggests a comprehensive assessment framework and modeling methodologies for evaluating system design alternatives that *simultaneously* serve different purposes. The review also reveals that integrating suitable technologies, business models, and long-term planning efforts offers significant synergistic benefits.

***Keywords:*** urban logistics; extreme event; essential supply; demand-responsive; autonomous vehicle; human-machine

# 1 Introduction

The growing concerns from both natural and anthropogenically induced disruptive events in recent years, as highlighted by Calvin et al. (2023) and Wannous and Velasquez (2017) have brought public and academic attention on resilience (Alimonti and Mariani, 2023; Cutter, 2021; Norris et al., 2008). Therefore, it is of interest to develop sustainable humanitarian supply chains to enhance infrastructure and community resilience (Kunz and Gold, 2017). In this paper, such a resilience refers to the ability of urban systems and communities to withstand, adapt, and recover from disruptions like natural disasters, technological failures, or other abrupt events such as military conflicts (Kirmayer and Whitley, 2009; Petit et al., 2013). Improved resilience is evident in reinforced transportation networks for extreme weather, dynamic smart energy grids, and rapid-response community support structures. Readers are referred to literature such as Chang et al. (2014) on general improvement in urban resilience.

In the meantime, there is a growing deployment and usage of transportation services that utilize readily available vehicles and the associated facilities to fulfill user mobility service requests in or near real-time (Mustapha et al., 2024). We refer to such services as mobility-on-demand (MOD) services. Unlike traditional public transportation with fixed routes and schedules, MOD services offer greater flexibility and convenience (Acheampong et al., 2020; Nam et al., 2018). These services can be understood as a function that matches user demand for transportation with available vehicles through a digital platform. This on-demand nature allows MOD services to adapt and respond dynamically to changing needs, making them a valuable tool for enhancing urban resilience. As urban environments continue to evolve, the environmental impacts of MOD services are drawing increasing research attention (Bahrami et al., 2022; Gurumurthy and Kockelman, 2022). Not only does it serve as a vital component of passenger transportation (Ma et al. 2019), but it also provides new levers for improving urban logistics (Snoeck et al. 2023; Ke et al. 2024). The rise of MOD services is also instrumental in facilitating automation, energy transitions, and the connectivity of future mobility systems, particularly in accelerating the adoption of connected autonomous vehicle (AV) and electric vehicle (EV) technologies and the usage of microtransit such as bike-sharing and scooter-sharing services. Additionally, as on-demand transit and aerial vehicles (e.g., Unmanned Aerial Vehicles (UAVs)) have been used for air taxis, package delivery, energy security, and communication systems enhancement in recent years, we consider them within the scope of the present paper. paper scope.

MOD services, characterized by their flexibility and adaptability, are a natural candidate for enhancing infrastructure and community resilience. They offer promising solutions in managing emergency scenarios, supporting evacuation processes (with higher occupancy), and ensuring the continuity of transportation services during critical periods. MOD services, including ride-hailing, customized buses, and shared e-bikes, may significantly enhance this resilience. In emergencies, they offer critical transport alternatives, facilitating evacuation and essential service access. Integrated with smart city technologies, these services contribute to efficient energy use and reduced congestion, aiding sustainable urban development (Yu et al. 2023). Their flexibility makes them key in emergency strategies, providing transport for essential personnel and supplies. Additionally, pre-negotiated contracts between public agencies and on-demand fleet companies can ensure swift mobilization during disasters, exemplifying a proactive approach to leveraging these services for urban resilience (Yu et al. 2023).

MOD-R services involve integrating available resources from different service providers to manage disruptions and maintain the system's resilience. The concept of "resilience as a service" (RaaS) has emerged in the field of operations management, where the goal is to reduce the recovery time during disruptions (Amghar et al., 2024). Resilience in urban transportation systems is important for maintaining the system's functionality in the face of disturbances and disruptions (Gonçalves and Ribeiro, 2020). Evaluating the resilience of urban traffic can help identify weak sectors and formulate response plans (Li et al., 2022). Resilience engineering for networked systems, including communication networks and the Internet, is an area that requires more explicit interest and research (Hutchison et al., 2023). Overall, implementing MOD-R involves integrating resources, evaluating resilience, and addressing disruptions to maintain system functionality. Figure 1 illustrates the concepts of using MOD services for improving infrastructure and community resilience.

Despite the growing importance and relevance of using MOD services for improving infrastructure and community resilience (here we refer to such services as MOD-R services), there lacks a comprehensive, forward-looking literature review. The review by Shah et al. (2023) and Qiang and McKenzie (2024) focus on using the use of mobility services (including MOD services) for improving public health. Dhall and Dhongdi (2023) review the use cases of UAVs for disaster monitoring, while (McDonald, 2019) focuses on using UAVs for stormwater management.

To address this gap, this paper presents a comprehensive literature review on this topic by delving into the current state of research on using on-demand mobility services for improving the adaptiveness of critical networks of urban transportation, electric power, and communication, in response to disruptive events. On-demand services that are dedicated solely or primarily to responding to disruptive events such as ambulances (Maghfiroh et al. 2018) and firetrucks are not considered in this paper.

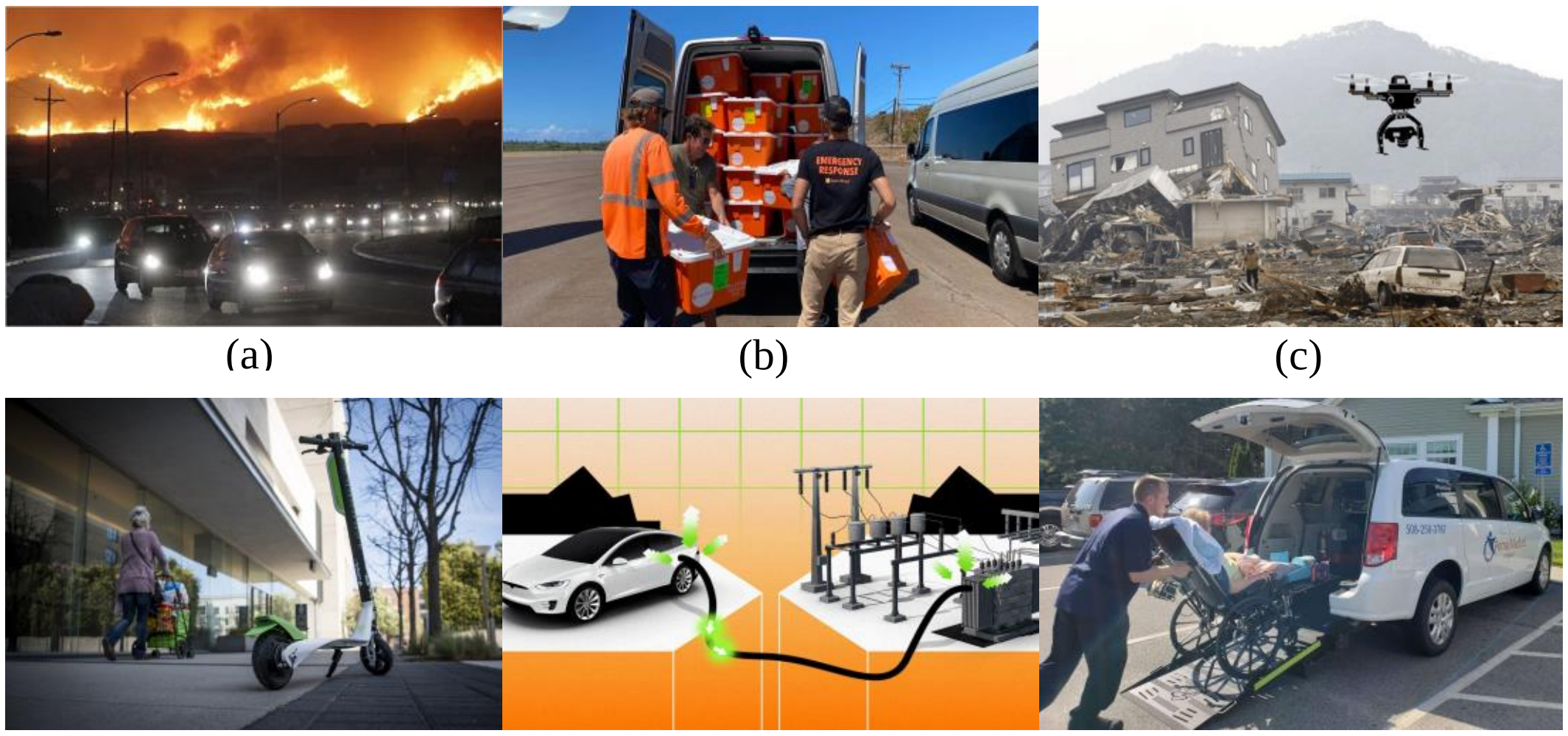

(a) (b) (c)

(d) (e) (f)

**Figure 1.** MOD services for (a) high-occupancy evacuation in wildfire (Fire Safe Marin, 2017), (b) essential supplies to vulnerable households (Direct Relief, 2023), (c) UAV probing, detection, essential supply delivery after earthquakes (Editor-in-Chief of Droneblog, 2023), (d) microtransit during pandemics (Bliss, 2020), (e) vehicle-to-grid (V2G) technology during power shut-offs and infrastructure repair (Donnelly, 2022), (f) urgent and non-urgent medical response (Prime Medical Transport, 2023).

The systematic literature review presented in this paper shows four main themes:

1. Resilient MOD services (R-MOD)
2. Empirical (retrospective) impact analysis and evaluation
3. Conceptualization and feasibility analysis for novel MOD-R usage
4. Enabling and automating technologies (e.g., electrification, automation, computing, and advanced communication)

The first theme, to an extent, sets a precursor and foundation for MOD-R, since if MOD services are not resilient, it is challenging to provide reliable MOD-R services. For example, if a storm forces the shutdown of MOD services, disaster response operators cannot use them for emergency supply delivery during the storm anymore. Although MOD-R has only recently sparked a strong interest, empirical analysis and system-level evaluations already exist as the second theme (Li et al., 2022). Papers in this theme examine and evaluate the impact from different perspectives, ranging from overall system performance, equity impact, and subjective scoring/comments from the survivors or disaster victims. The third theme – ideation and benefit-cost analysis MOD-R – is of core interest to the present paper. In this theme, we find a wide range of applications of MOD-R services and foresee even more novel usages to emerge. The papers in the fourth theme propose enabling and augmenting technologies, such as ad hoc communication networks during disasters, explicitly for MOD-R. Electrification, automation, and advanced communication technologies can significantly increase the ability and scope of MOD-R services in the areas of anomaly detection, evacuation, grid stabilization, essential supplies, among others. Additionally, the literature review also reveals the importance of human roles in enabling and enhancing MOD-R services, indicating the need for improving human labor training and facilitating equipment and user-friendly decision-support systems.

Combining these findings, three main areas of research gaps emerge –

1. the need of effective integration among technologies,
2. the need of utilizing joint power of human intelligence and artificial intelligence for planning and operating MOD-R services, and
3. the need to embed MOD-R services into the overall long-range transportation and community planning.

Collectively, this paper refers to them as the needs of synergistic design, development, and operation of MOD-services.

The present paper contributes to the literature in three folds. **First,** the paper presents the first systematic review of MOD-R related literature. Based on the review, we classify existing papers into four main themes and provide an in-depth discussion on each theme and across themes. **Second,** the paper introduces a framework to address the need of proactive design and development of MOD-R systems to maximize the synergistic effects between MOD services and technologies such as automation, electrification, advanced sensing data processing, cybersecurity, and advanced communication. **Lastly,** due to the tremendous challenges of fully automating the MOD-R services anytime soon, the framework also highlights the promising collaborations between intelligent vehicles and humans (e.g., residents, social workers, disaster response operators, emergency medical technicians, social workers, repair crew, emergency planners) for facilitating the transition phases in the decades to come, which has strong practical implications for MOD service providers and disaster relief organizations.

The remainder of this paper is structured to provide the methodology of the literature review and an exploratory analysis of the publication trends and keywords. Through an in-depth analysis by theme, we aim to offer a holistic view of the current landscape and future possibilities, setting the stage for informed discussions and strategic planning of utilizing MOD-R services as an integral component of urban resilience measures. Lastly, we envision a more coordinated and streamlined collaboration between human workers and vehicles for enabling and enhancing MOD-R services.

# 2 Methodology

In conducting this literature review, our methodology was designed to systematically explore the intersection of MOD services and their role in enhancing urban infrastructure and community resilience. The primary databases for the literature search were Scopus and Google Scholar. The first step of the literature review involved defining the research questions that would guide the entire process. These questions were centered around the business models of the MOD-R services (Service Model), their means of transporting people and resources (Transport Tool), the key objectives of the MOD-R services (Purpose), and the specific disruptive events (Scenario). The search strategy was crafted, utilizing a range of terms relevant to our research questions in terms of Service Model, Transport Tool, Purpose, and Scenario. These keywords are connected using the logic of "AND." Within each set, the terms are associated using the logic of "OR." The terms are presented in Table 1. The scope of the search was further confined to journal and conference articles written in English since 1960. These specific terms are obtained through an iterative process. This initial identification process reveals 667 articles.

The screening phase first drops duplicates, leaving 665 articles. Then, the titles, abstracts, and keywords are reviewed to further exclude the papers that are not focused on MOD-R. For example, Fezi (2020) provides a broad overview on the potential of architecture and urbanism in the prevention and control of epidemics and the improvement of public health, where shared mobility and robo-taxis are only briefly mentioned as examples; hence, this paper is not included in the in-depth review.

The screened articles formed the next crucial phase. The articles were identified through our search strategy. A systematic screening process was also employed, where articles were first filtered based on their titles, abstracts, and keywords to gauge their relevance to our research objectives. This was followed by a thorough review of the full texts of these preliminarily selected articles. To enrich the review further, the references within these articles were examined to uncover any additional pertinent literature. To maintain the focus and integrity of our review, the process established specific inclusion and exclusion criteria. Articles were selected based on their direct relevance to the themes of on-demand mobility and urban infrastructure resilience. Key criteria for inclusion encompassed discussions around urban transportation systems, the role of MOD services in infrastructure resilience, and insights into the human roles in MOD-R services. This methodological approach ensured a comprehensive and systematic exploration of the literature, laying a solid foundation for the subsequent analysis and discussion of our findings in the results and discussion sections of the paper.

**Table 1.** Search term combinations

| **Service Model** | **Transport Tool** | **Purpose** | **Scenario** |
|---|---|---|---|
| Shared Mobility, On-Demand, Demand-Responsive, Ride-hailing, Ride-sharing, Car-sharing, Vehicle-sharing | Vehicle, Car, Scooter, Bike, Bicycle, UAV, Taxi, Drone, Ferry, Boat, Truck, Bus, Shuttle | Resilience, Resilient, Adaptive, Adaptable, Rescue, Evacuate, On-Site Care, Relocate, Repair, Disaster Relief | Disaster, Disruptive Events, Disruption, Extreme Events, Emergency, Hurricane, Earthquake, Seismic, Heat, Wildfire, Storm, Blizzard, Storm. Pandemics, Flooding, Landslide |

As suggested in the introduction section, the review process reveals four main themes of MOD-R related papers. The details of how these four themes emerge will be presented in Section 3. Section 4 will provide more in-depth analysis of these themes. Also note that, for the first theme, we specifically focus on resilient MOD in response to major disruptive events. Existing literature on MOD in response to (common or recurrent) demand fluctuation is not the focus of this theme; instead, this type of literature would be considered *reliable* MOD by the present paper, and hence, not included in the review. Figure 2 shows the screening process and the results in terms of the number of papers filtered at each step.

Once the papers are selected, we use mapping (e.g., word map, literature co-occurring map, and correlation map) to help identify literature patterns. As will be further explained in Section 3, this process reveals four major themes as forementioned. Another important theme that is present in all four themes, is the consideration of human roles in enabling or supplementing MOD-R services, which we have left for a forward-looking discussion. The in-depth content analysis also reveals two main research gaps: (1) lack of studies on systematic integration among technologies, and (2) lack of consideration for the endogenous/interconnection between MOD-R development and the natural and built environments. Combining these two research gaps with the importance of human-roles, this paper collectively refers to them as the research gap of synergistic design and development. This process is illustrated in Figure 3.

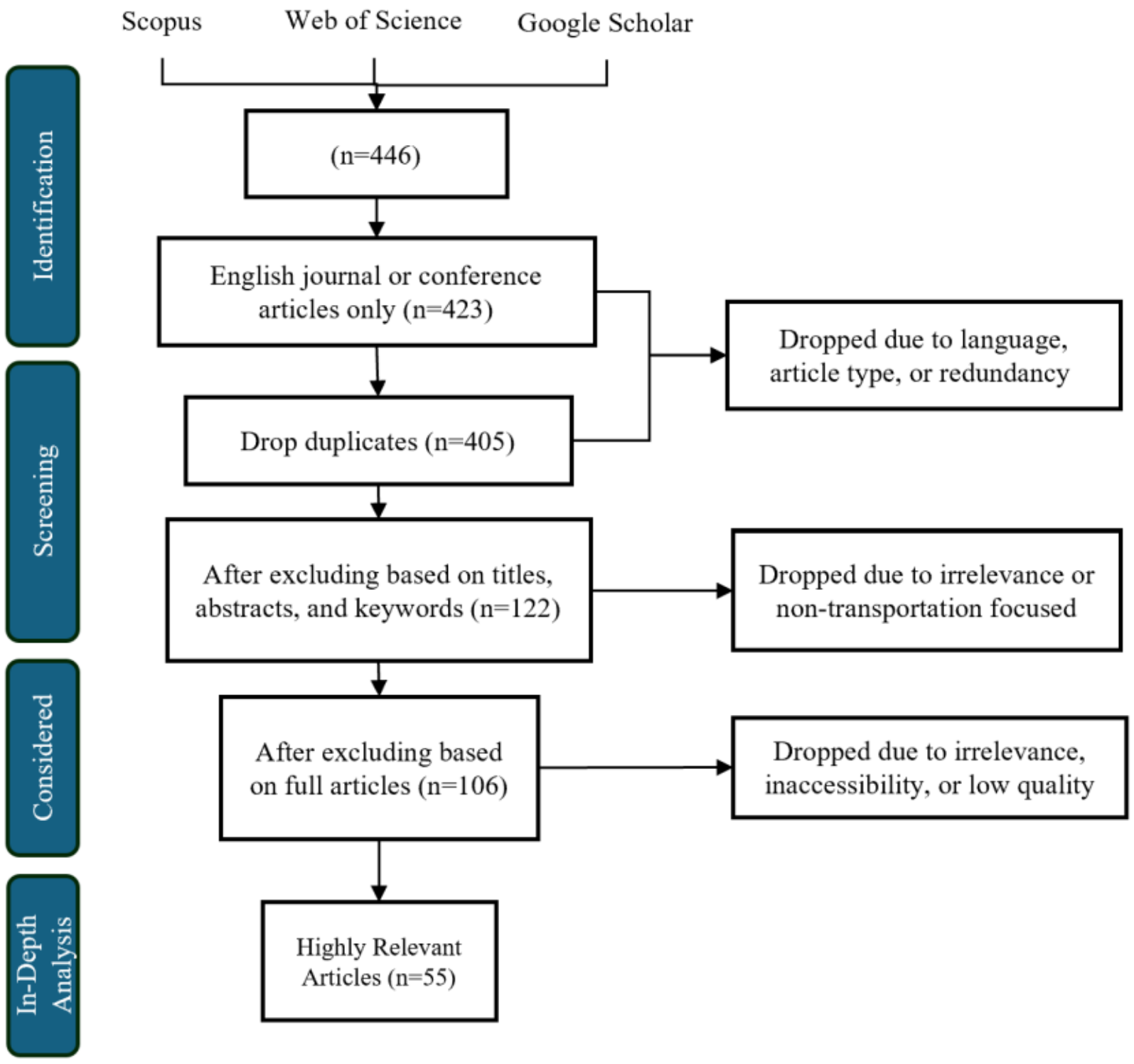

**Figure 2.** Article selection process and result

Bibliometric networks are the way bibliometric research is visualized. A bibliometric network consists of nodes and edges, where nodes can be publications, journals, researchers, or keywords, and the edges represent the relations between pairs of nodes. Among the different types of available bibliometric networks, this study uses the co-occurrence of author keywords as well as index keywords (terms extracted from titles and abstracts), which allows for the exploration of relationships among topics in a research field and therefore supports the identification of existing patterns and potential research gaps. VOSviewer (van Eck and Waltman, 2017) is the tool to construct and visualize bibliometric networks in this paper. This paper will use two types of visualization: network visualization and density visualization. Santamaria-Ariza et al. (2023) and Moosavi et al. (2022) are example review papers that utilize the power of this tool. Facilitated by VOSviewer and other analytical methods such as word map and correlation heat map, the included articles (from the last step of the screening process illustrated in Figure 2 are further categorized into multiple themes and research gaps. This process is illustrated in Figure 3. The details of this process will be shown in Section 3, 4, and 5.

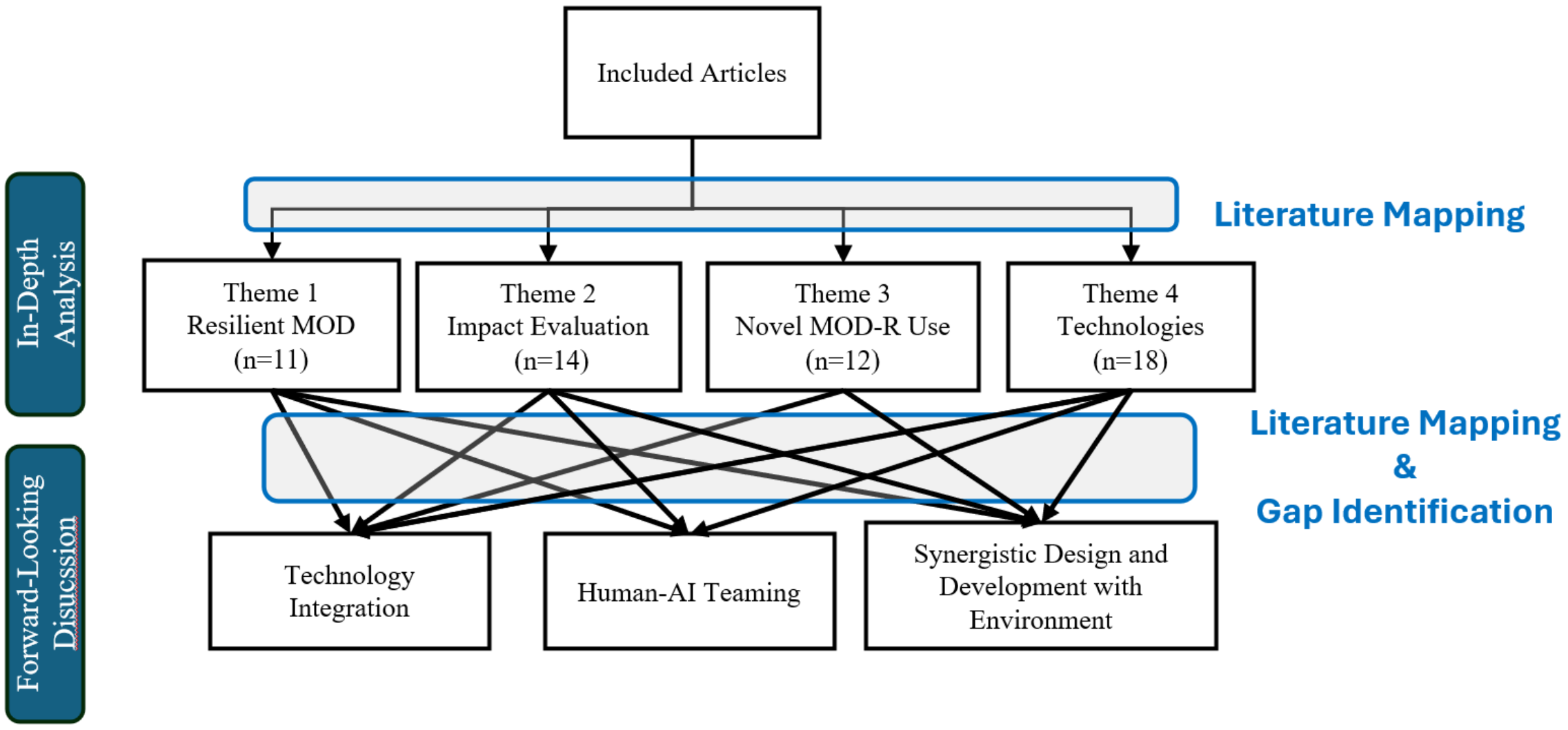

**Figure 3.** Further classification and insight development from the highly relevant literature

# 3 Exploratory Analysis

This section presents the results of exploratory analysis of MOD-R related literature. The analysis aims to uncover thematic focuses, key trends, and prevalent topic connections in this research area. We first present the reasoning behind the emergence of the four themes. Then, we present the trend analysis. The analysis aims to uncover key trends, thematic focuses, and prevalent keywords in this research area.

The word map (Figure 4) visualizes the frequency of author keywords within the reviewed literature. It provides a snapshot of the most prominent topics and concepts in MOD-R related articles. This map is instrumental in identifying the central themes and terminologies that dominate the discourse. The word map reveals two general categories of keywords – (1) design, management, and evaluation of MOD-R (e.g., “management,” “demand,” “analysis”), and (2) technologies (e.g., “communication,” “antenna,” “computing,” “ad hoc network”). To further categorize the literature, we next use literature maps.

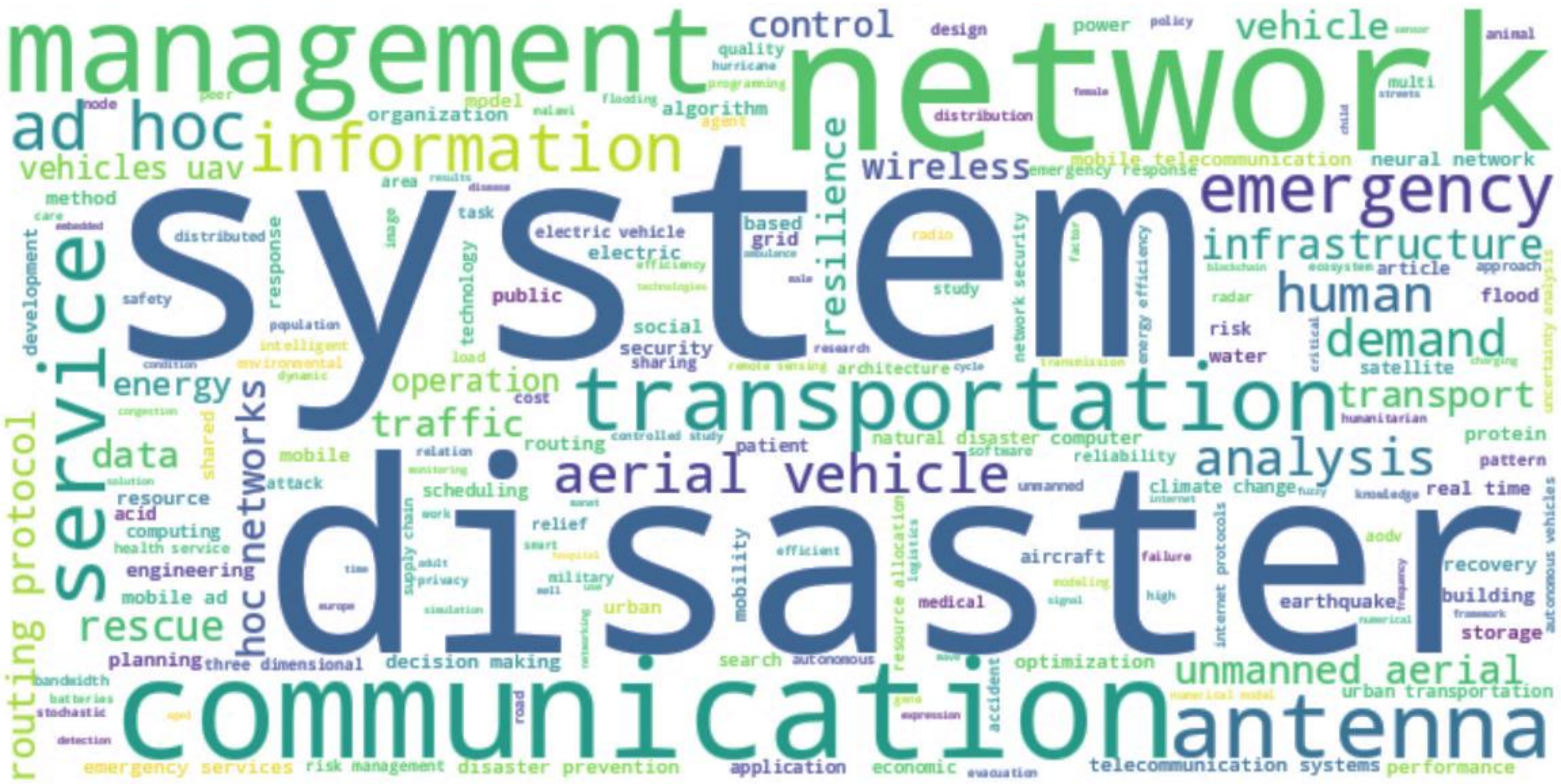

**Figure 4.** Word map based on the author keywords

Figure 5 shows a literature co-occurring map of the post-screened literature. We see that the map confirms the existence of the topics related to the “technologies” category (in blue). In addition to this category, the map further splits the second category “design, management, and evaluation” into four main subcategories: empirical impact evaluation of MOD-R (in red), Resilient MOD (in green), Novel MOD-R Usage (in yellow), and human roles (in purple). After a careful in-depth content analysis, it is found that the literature associated with the first subcategories have relatively stand-alone papers. The papers related to the fourth subcategories, “human-roles,” generally also fall into the “empirical impact evaluation of MOD-R” subcategory, the “resilient MOD (R-MOD)” subcategory, and the “technologies” subcategory. Therefore, in Section 4, we first examine in-depth the following four categories (the “technology” categories plus the first three “subcategories”). During the content analysis, a key research gap is identified, where there lack comprehensive studies of how to integrate different mature and emerging technologies to maximize the synergies for MOD-R purposes and how these technologies interact (rather than only affect) the natural and built environments. Combining this insight with the subcategory “human roles”, Section 5 is added to address two types of synergies – (1) the synergies among technologies, (2) the synergies between human intelligence and artificial intelligence, and (3) the synergistic design and development that take into account the interactions with the natural and urban environment.

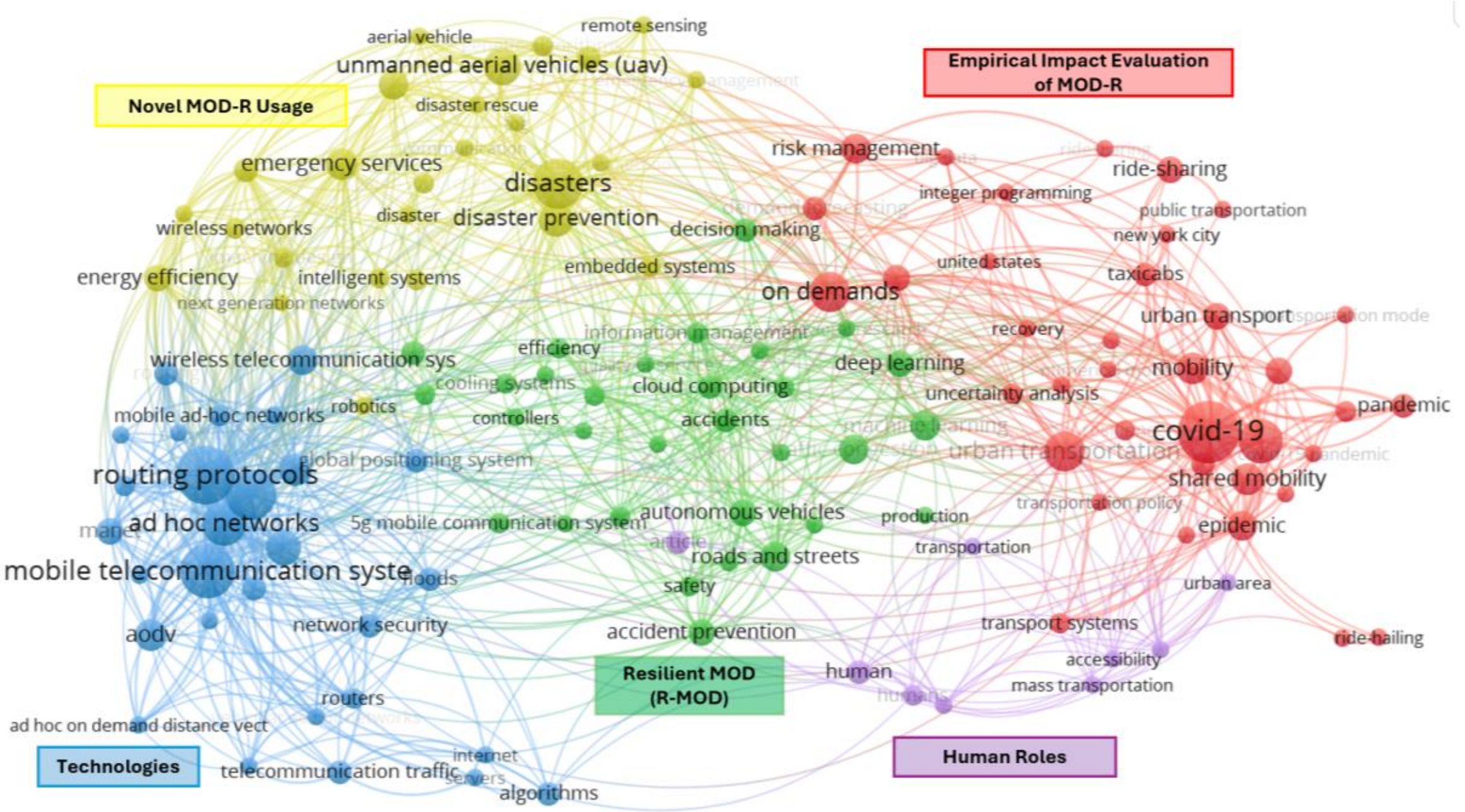

**Figure 5**. Literature co-occurring map for the literature.

The stacked trend map (Figure 6) offers a comprehensive view of the evolution of research themes over time, up to Year 2023. Each layer in the map represents a different theme, including Resilient MOD Services, Conceptualization and Feasibility, Enabling and Automating Technologies, and Empirical Impact Analysis. The map illustrates the growth of literature in each area, highlighting the increasing emphasis on MOD-R related services in recent years. It is evident that Conceptualization and Feasibility, along with Enabling and Automating Technologies, have garnered significant attention, indicating a strong research focus on the foundational aspects of MOD services. An interesting pattern is a noticeable surge from Year 2013 to Year 2016, which seems to be related to the public perception of a seemingly increased frequency and severity of natural and society-induced disasters such as the tsunami-induced Fukushima nuclear accident in 2011 and Hurricane Sandy that ravaged the Caribbean and the coastal Mid-Atlantic region of the United States in 2012. Then, there is a minor decline in the published articles. A new surge from 2020 was likely triggered by the global pandemic and the maturing technologies for electrification, automation, and advanced communication.

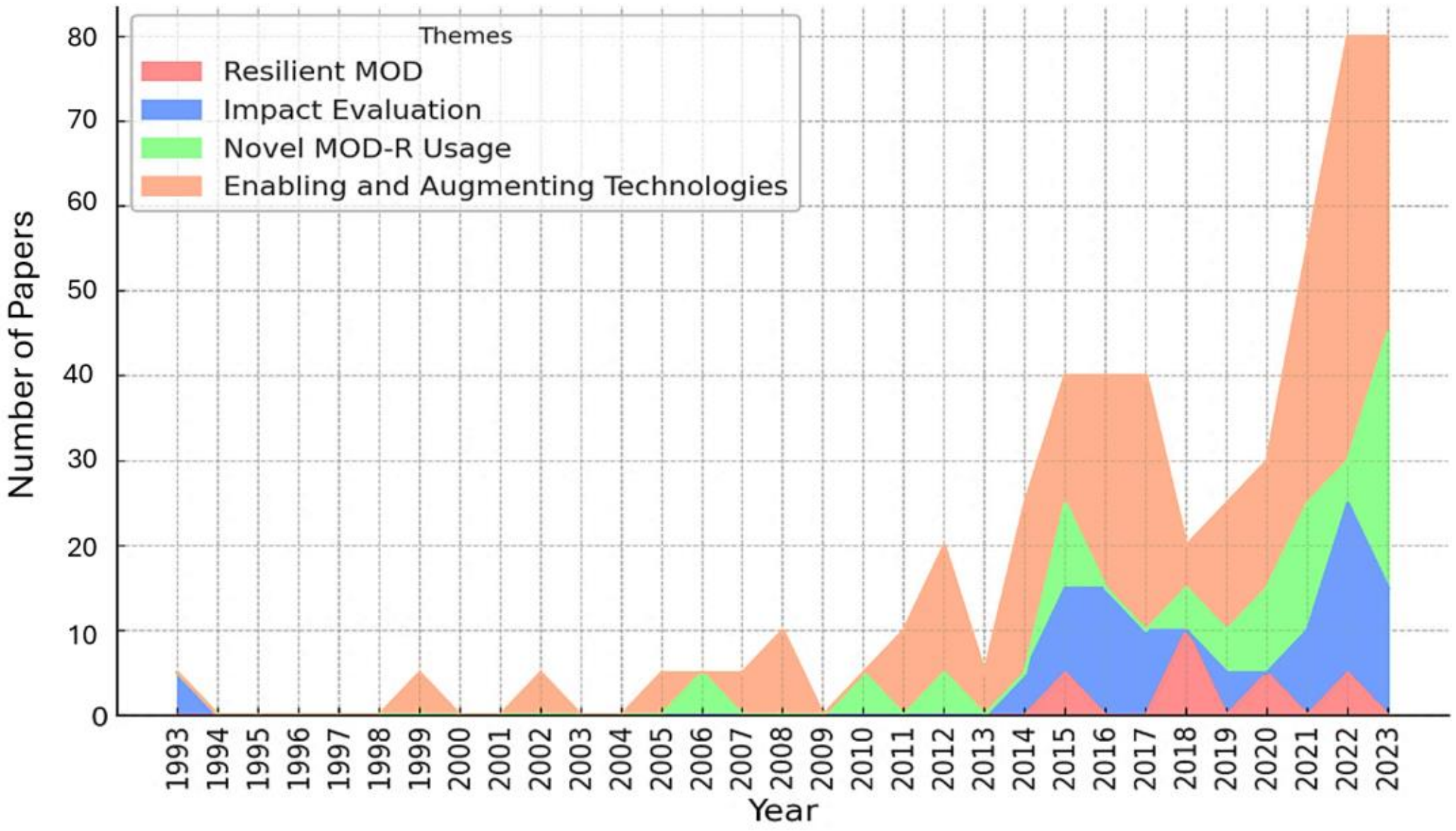

**Figure 6.** Stacked (non-cumulative) trend plot of the MOD-R related articles by year

The exploratory analysis of the literature related to MOD-R services reveals a dynamic and rapidly evolving field that is strongly affected by (then) recent events and technological advancement. The analysis also shows that the field is highly interdisciplinary and presents a clear picture of the research landscape, highlighting the growing importance of MOD-R services. The convergence of technology and mobility underscores the potential of MOD services in shaping future urban infrastructure and community resilience strategies.

# 4 In-Depth Content Review

This subsection provides an in-depth discussion of the papers by theme, including their commonalities, differences, and research gaps. As Theme 2 and 3 are considered the most directly related to the present paper, they are referred to as the "Core" papers. To provide an overall and intuitive understanding of the four themes, Figure 7 visualizes the keyword densities for the selected papers in each theme. The size of the keywords (and the heat level) indicates the frequency of the keywords appearing in the corresponding theme of literature. The shorter (longer) distance between two keywords indicates that the keywords are more (less) likely to appear in the same article. We can see that Theme 1 concerns greatly with public transit and policy support. Papers in Theme 2 tend to utilize data during and after the COVID 2019 pandemic for impact analysis. Theme 3 shows that proposing new MOD-R uses are often associated with three main groups of concepts – communication infrastructure and equipment (e.g., antennas), modeling and evaluation methods (due to the hypothetical nature of the proposals), and built environment (e.g., buildings and housings). The density visualization also shows that Theme 4 focuses greatly on communication technologies and (unmanned) aerial vehicles. Table 2 provides a summary of selected papers from a different perspective. The table further selected representative papers from the "Core" papers in terms of the types of disruptive events, specific MOD-R purposes, augmenting technologies, and study methods. These selected papers are also representative as the discussed MOD-R services often require their own resilience (Theme 1) and demand enabling technologies such as advanced communication, automation, electrification, and advanced decision support systems and algorithms (Theme 4).

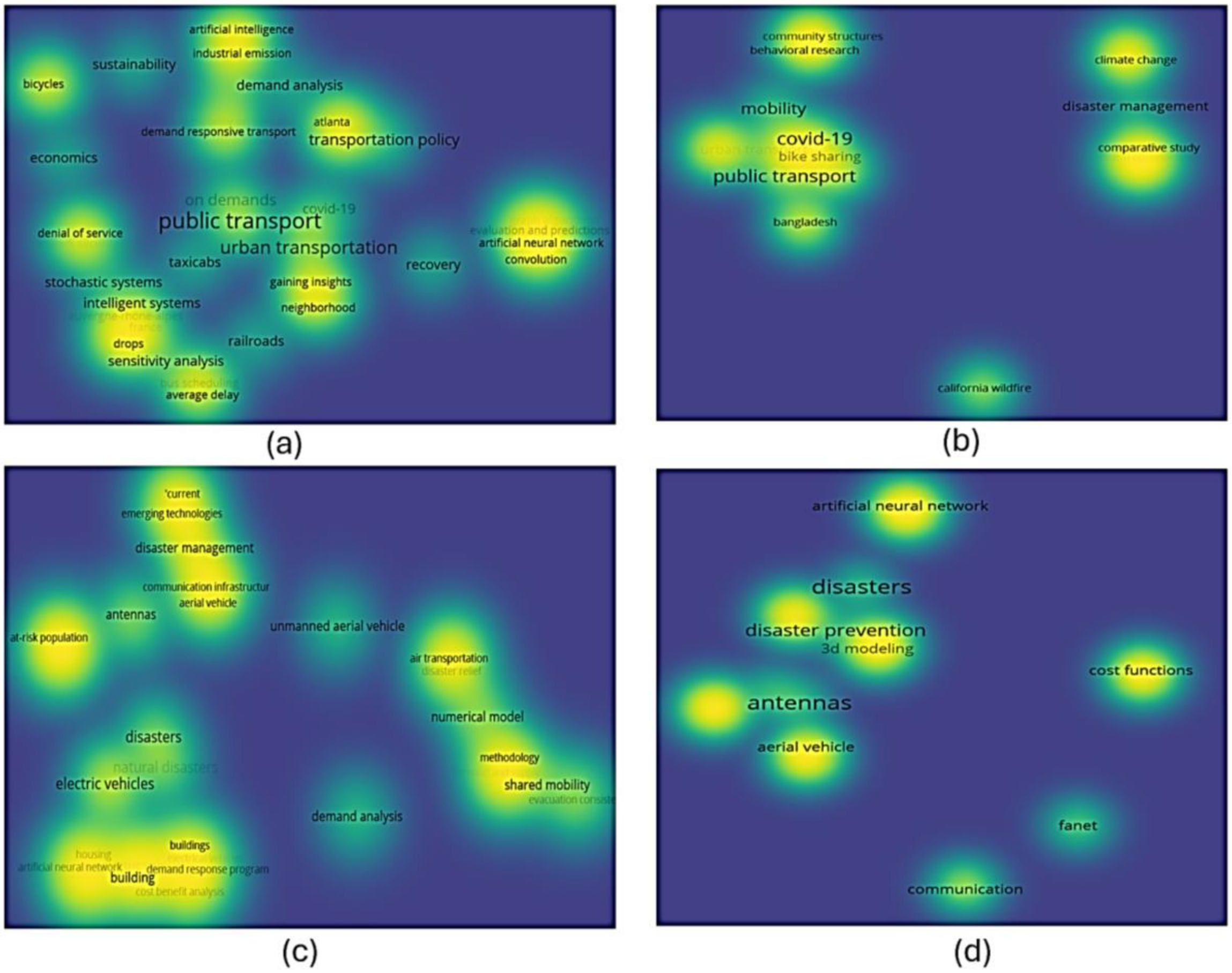

**Figure 7. (a) Density visualizations of Theme 1 papers. (b) Density visualizations of Theme 2 papers. (c) Density visualizations of Theme 3 papers. (d) Density visualizations of Theme 4 papers**

## 4.1 Resilient MOD Services (R-MOD) (Theme 1)

R-MOD, in a sense, is a precursor and foundation of (reliable) MOD-R. The papers in this theme highlight a transformative approach to enhancing urban transportation resilience through strategic infrastructural integrations, technological innovations, and flexible service models. This synthesis delves into the multifaceted dimensions of R-MOD by collating insights from a spectrum of studies that explore the practical and theoretical implications of enhancing urban transportation systems' adaptability and responsiveness to disruptions.

Regarding strategic infrastructure and technological integration, the studies by Henry et al. (2022) and (Wang et al. 2020) represent a significant convergence of computational modeling and strategic infrastructure deployment aimed at mitigating transportation disruptions. Henry et al. (2022) emphasize the importance of location optimization for park-and-ride facilities to enhance network resilience under varied disruptive scenarios. They employ a stochastic approach to capture user behavior and optimize facility locations, which significantly contributes to maintaining service continuity amidst recurrent urban disruptions like adverse weather or labor strikes. Similarly, Wang et al. (2020) employ a novel diffusion graph convolutional approach to model and predict transportation resilience under extreme weather conditions, highlighting how deep learning can elucidate the spatiotemporal patterns of resilience and improve response strategies. The dimension of collaborative governance in enhancing transportation resilience is explored by Ma et al. (2018), who analyze the governance challenges posed by free-floating bike-sharing systems in Shanghai. They highlight the necessity of integrating emerging social actors and adjusting governance models to facilitate effective collaboration among government, business, and societal stakeholders, thus ensuring the sustainability and resilience of shared mobility services.

Regarding operations, Campisi et al. (2021) and Liyanage et al. (2019) underscore the critical role of flexible mobility solutions, particularly under changing circumstances such as the COVID-19 pandemic. Campisi et al. (2021) reflect on the accelerated shift towards Mobility-as-a-Service (MaaS) platforms in Italy, suggesting that demand-responsive transport (DRT) systems can effectively reduce urban congestion and environmental impact while enhancing resilience by adapting to fluctuating user demands. Liyanage et al. (2019) discuss the integration of on-demand shared mobility with established public transport frameworks, leveraging AI, IoT, and cloud computing to foster a seamless and responsive urban transport system. The operational challenges and adaptations during the COVID-19 pandemic have been critically examined by Auad et al. (2021), who study the resilience of On-Demand Multimodal Transit Systems (ODMTS) in Atlanta. Their research reveals how integrated networks of high-frequency transit options coupled with on-demand services can offer resilient and cost-effective solutions during various stages of the pandemic, ensuring continuity even with depressed demand and stringent public health protocols. Addressing the cybersecurity aspects of resilient mobility services, Thai et al. (2018) provide a theoretical and practical framework for understanding and mitigating risks associated with denial-of-service attacks on MaaS systems. Their approach combines queuing theory and economic strategies to safeguard system availability, emphasizing the need for robust security measures to protect the integrity of on-demand mobility services.

These studies collectively illuminate the complex interplay between infrastructure planning, technological integration, service flexibility, and governance in the context of R-MOD. Despite the robust theoretical frameworks and innovative approaches presented, a noticeable gap remains in the empirical validation of these models across diverse and multi-scale disruptive events. The role of human decision-makers and the impact of policy frameworks in operationalizing these resilient services during crises are areas that require further exploration. Future research should focus on empirical studies that test these frameworks in varied real-world scenarios, thereby providing a more comprehensive understanding of their effectiveness and informing the development of more resilient urban transportation systems.

## 4.2 Empirical Impact Analysis and Evaluation of MOD-R (Theme 2)

The second theme in the systematic literature review on MOD-R services focuses on empirical impact analysis, highlighting real-world evaluations and adaptations of logistics modalities during disruptive events. The papers within this theme explore the resilience and adaptability of various mobility services, especially during the COVID-19 pandemic, which has significantly altered transportation dynamics globally. This synthesis will integrate findings from several studies to outline key observations and insights into the operational impacts and user responses to transportation disruptions.

MOD-R's role on for improving (other) transportation systems components, which in turn contributes to the overall transportation infrastructure resilience and broader society. Borowski et al. (2023) investigated how MOD-R services respond to unplanned rail disruptions in Chicago. Their analysis revealed that ridesourcing, an instance of MOD services, could provide adaptive capacity during transit disruptions, although its benefits were not equitably distributed across different community demographics. This study points to the potential of ridesourcing to complement existing transit networks during disruptions, suggesting a need for policies that address equity issues in access to these adaptive mobility services. Existing research has also examined the impact of bike-sharing systems on community resilience, Teixeira et al. (2022) and Qin and Karimi (2023) provide valuable insights into the role of bike-sharing systems on community resilience during the COVID-19 pandemic. Teixeira et al. (2022) found that despite a decline in overall usage due to lockdowns and teleworking, bike-sharing in Lisbon continued to serve essential travel needs with perceived lower infection risks compared to public transport. The study suggests that the inherent flexibility and perceived safety of bike-sharing can enhance urban transport resilience by providing a reliable alternative during public health crises. Qin and Karimi (2023) examine the spatiotemporal dynamics of bike-sharing in Pittsburgh, noting that the service demonstrated resilience by adapting to changing usage patterns throughout the pandemic. These findings underscore the role of bike-sharing in maintaining urban mobility amidst disruptions, warranting policy support such as expanded infrastructure and integrated service models.

The role of trust during disasters plays a critical on the effectiveness of resource (including mobility resource) sharing. Wong et al. (2021), Idziorek et al. (2023), and Islam et al. 2023), delve into the social and inter-organizational dynamics of resource sharing during emergencies, specifically exploring the influence of trust and compassion on sharing behaviors. Wong et al. (2021) document the willingness to share mobility and sheltering resources during the California wildfires, finding that higher trust and compassion levels significantly enhance sharing behaviors. This insight is critical for developing community-based strategies to foster resilience through social capital. Idziorek et al. (2023) extend this analysis to a cross-cultural context, comparing U.S. and Japanese communities' willingness to share essential resources post-disaster. They concluded that strong social ties and trust are pivotal in mobilizing community resource sharing, highlighting the need for trust-building interventions in disaster preparedness. Islam et al. (2023) examine the obstacles of vehicle sharing among non-governmental organizations during disaster response operations through interviewing experts. They found five main groups of barriers that linked with many potential side-effects or uncertainties such as accountability for possible vehicle damage and accidents, indicating the need of external (“third party”) facilitators.

The investigation of the long-term shifts in travel preferences is critical for understanding the effectiveness and influence of MOD-R services. The papers by Mahmud et al. (2024) and Chang et al. (2024) address the long-term (month-by-month or year-by-year) behavioral changes in travel preferences due to the pandemic. Mahmud et al. (2024) reported a shift towards private cars and active transportation in Dhaka, influenced by public perceptions of safety and changes in lifestyle, such as increased teleworking and use of delivery services. Chang et al. (2024) utilized a graph-based deep learning approach to analyze COVID-induced changes in travel mobility, revealing significant shifts in travel patterns and network structures. These studies highlight the need for adaptive urban transport solutions that can accommodate evolving travel behaviors.

To further understand the connections between papers in Theme 1 and Theme 2, Figure 8 shows the co-occurrence literature network using the (author and index) keywords from papers in both themes. Jointly considering Figure 8 with Figure 7 (a) and (b), we see that public transportation and COVID-19 are two closely linked topics, surrounding which are spatiotemporal analysis under uncertainty (in red), machine learning and numerical models (in blue), multi-modal economics and efficiency (in yellow), and demand consideration and policy impacts (in green).

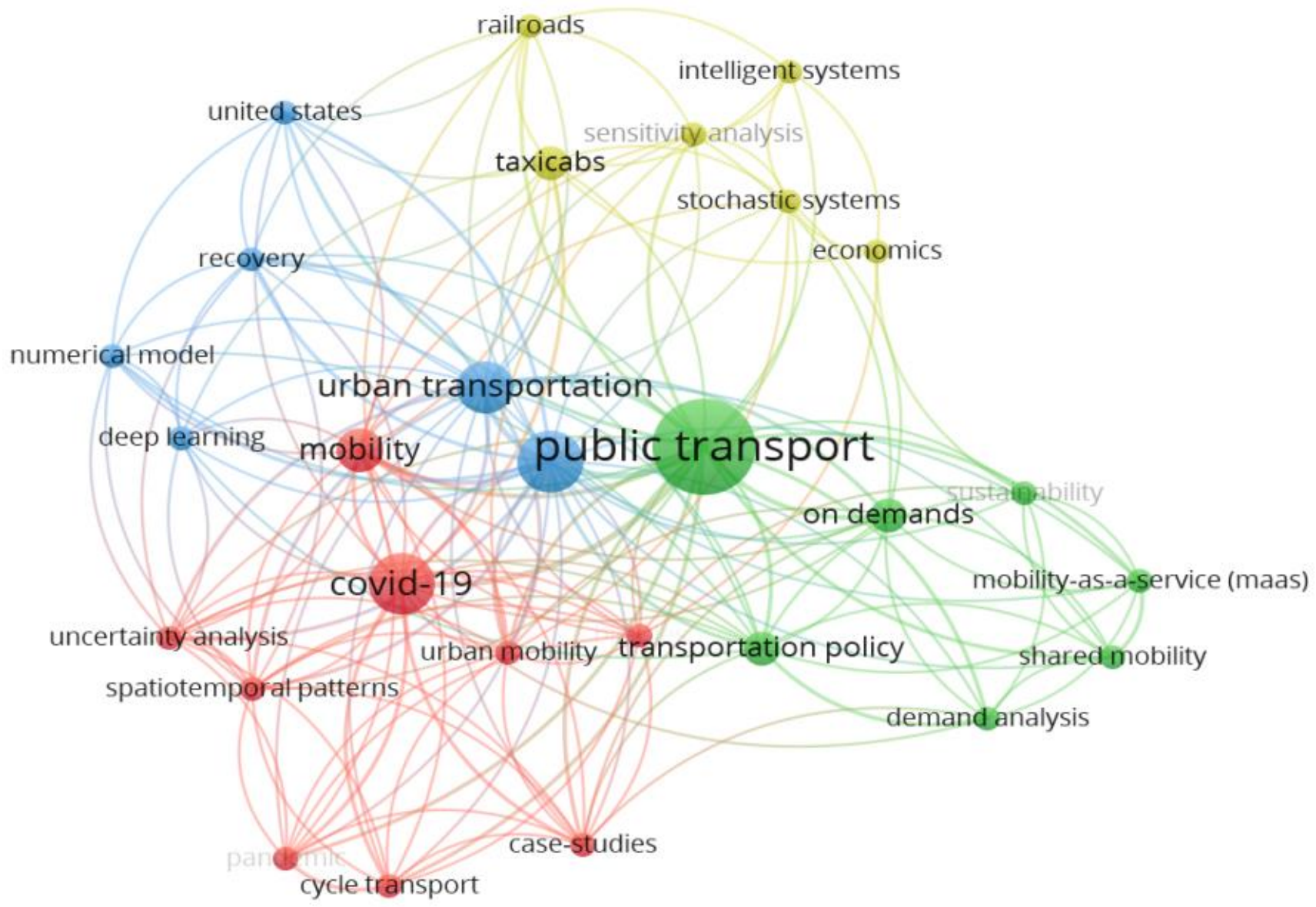

**Figure 8**. Co-occurance network of literature in Theme 1 and 2

Integrating these findings, it is evident that resilience in urban transportation is multifaceted, encompassing technological adaptability, social dynamics, and long-term behavioral changes. The empirical evidence from these studies highlights the critical role of on-demand transport systems like bike-sharing in maintaining mobility during disruptions and the foundational importance of social trust and community cohesion in enhancing system resilience. Moreover, the observed shifts in travel preferences and the adaptive responses of MOD-R services underscore the complexity of urban mobility resilience, which requires comprehensive and inclusive policy frameworks to support equitable and sustainable transportation ecosystems. These insights collectively emphasize the necessity of a nuanced understanding of transport resilience, where policy interventions should be informed by empirical evidence and tailored to address the diverse needs and behaviors of urban populations. Further research is needed to explore the interactions between different transportation modes and community responses to enhance the overall resilience of urban transport systems in the face of future disruptions. Furthermore, the psychological resilience of first responders and its impact on decision-making and effectiveness in crisis situations is an under-researched area that demands attention. Indeed, understanding the psychological and behavioral aspects of individuals, particularly travelers and disaster response operators under stress and pressure, is vital. Although Yu and Hyland (2020) propose a modeling method for individual decisions under pressure and stress for on-demand mobilities, the methods have not been empirically tested in extreme scenarios. Potential future research may involve exploring the psychological and behavioral aspects of individuals during emergencies, providing insights into how MOD-R services can be tailored to better support decision-making in high-pressure situations. Lastly, the interaction between MOD services and the broader emergency response systems requires further exploration to understand how MOD can be effectively utilized during crises and integrated into broader urban resilience strategies. Third, there is a need for empirical research that integrates technology, policy, operations, and human factors into a cohesive framework for urban resilience. This integration is crucial for validating and testing the efficacy of proposed resilience frameworks (Liu et al. 2021; Datola 2023; Ribeiro et al. 2019) in diverse scenarios. Future research should focus on developing standardized resilience metrics and benchmarking tools to evaluate the impact of MOD-R, necessitating cross-disciplinary collaboration.

## 4.3 Novel Usage and Incentives of MOD-R (Theme 3)

Articles in the third theme delve into the conceptualization and feasibility analysis of novel uses of MOD-R services, particularly in the context of disaster management and resilience planning. This theme explores innovative approaches to leverage on-demand mobility technologies, including ride-sourcing, unmanned aircraft systems (UAS), and shared autonomous electric vehicles (SAEVs), to enhance infrastructure and community resilience during various disruptive events. The papers in this theme present a range of models and strategies aimed at improving the effectiveness of MOD services in emergency and disaster situations. Note that earlier paper proposals of novel MOD-R use might later be implemented (and hence not novel anymore), but we still consider these earlier papers in this theme considering that the nature of these papers is to propose new ideas (relatively to the *then* MOD-R practices when the papers were released).

MOD-R has been proposed to be temporarily recruited for evaluation and rescue purposes. (Wang et al. (2021) address the potential of ride-sourcing services in hurricane evacuation scenarios, presenting mathematical models that balance demand and supply through dynamic pricing mechanisms. They propose a subsidy model to ensure that vulnerable populations can access these services, highlighting the role of ride-sourcing in augmenting public resources during evacuations. This study demonstrates the feasibility of integrating ride-sourcing into disaster response frameworks, potentially improving evacuation efficiency while addressing economic considerations. Ahmed et al. (2020) explore the influence of social networks on hurricane evacuation decisions and the capacity for shared evacuations. Using a personal network research design, they employ multinomial logistic regression and zero-inflated Poisson models to understand how day-to-day interactions and demographic similarities impact evacuation behaviors. Their findings underscore the importance of social dynamics in evacuation planning and suggest that leveraging existing social ties can enhance shared evacuation strategies.

The integration of multiple technologies and business models for a spectrum of resilience-related purposes is also worth highlighting. Yu et al. (2023) propose the simultaneous use of SAEVs to support disaster response efforts, including monitoring, evacuation, emergency supply delivery, and infrastructure repair. They present a modeling framework for the strategic deployment of SAEVs in various disaster scenarios, illustrating how these vehicles can enhance the adaptiveness and resilience of urban systems. Tian and Talebizadehsardari (2021) and Sridharan et al. (2023) both focus on building energy resilience through shared parking stations for electric vehicles. They explore energy management strategies that utilize electric vehicles in shared facilities to maintain power supply during outages, showcasing how peer-to-peer energy sharing can mitigate the impacts of power disruptions. Kim and Davidson (2015) summarize the applications of unmanned aircraft systems (UAS) in disaster management, discussing the regulatory, safety, and privacy challenges associated with their use. They advocate for collaborative efforts to integrate UAS into existing disaster response protocols, emphasizing the potential of UAS to perform critical tasks such as search and rescue, medical supply delivery, and infrastructure assessment. Jeong et al. (2020) introduce the concept of a humanitarian flying warehouse (HFW), an innovative solution using high-altitude airships and UAVs to deliver supplies to conflict zones safely. They validate their model through a case study, showing that the HFW can significantly enhance the safety and efficiency of delivering aid, reducing the risks associated with ground transport of humanitarian supplies. Kirubakaran and Hosek (2023) propose the use of tethered unmanned aerial vehicles (TUAVs) to maintain communication networks during disasters. By optimizing TUAV deployment with genetic algorithms, they demonstrate how these systems can provide reliable connectivity, essential for effective disaster response and management. Patil et al. (2022) suggests a wide range of uses of UAVs in rescue operations. They also highlight potential challenges, such as vicissitudes of wind flows, remote control connection issues, heavy rains, in effective deployment of UAVs.

To better understand the relationship of the papers in Theme 1 and Theme 3, Figure 9 shows the co-occurrence literature network. Jointly considering Figure 9 with Figure 7 (a) and (c), we see four main groups of concepts: supporting public transport under uncertainty (in red), disaster prevention and management (in yellow), electrification (in green), and demand consideration (in blue). Furthermore, demand consideration and public transportation under uncertainty are particularly associated and often appear in papers in both themes.

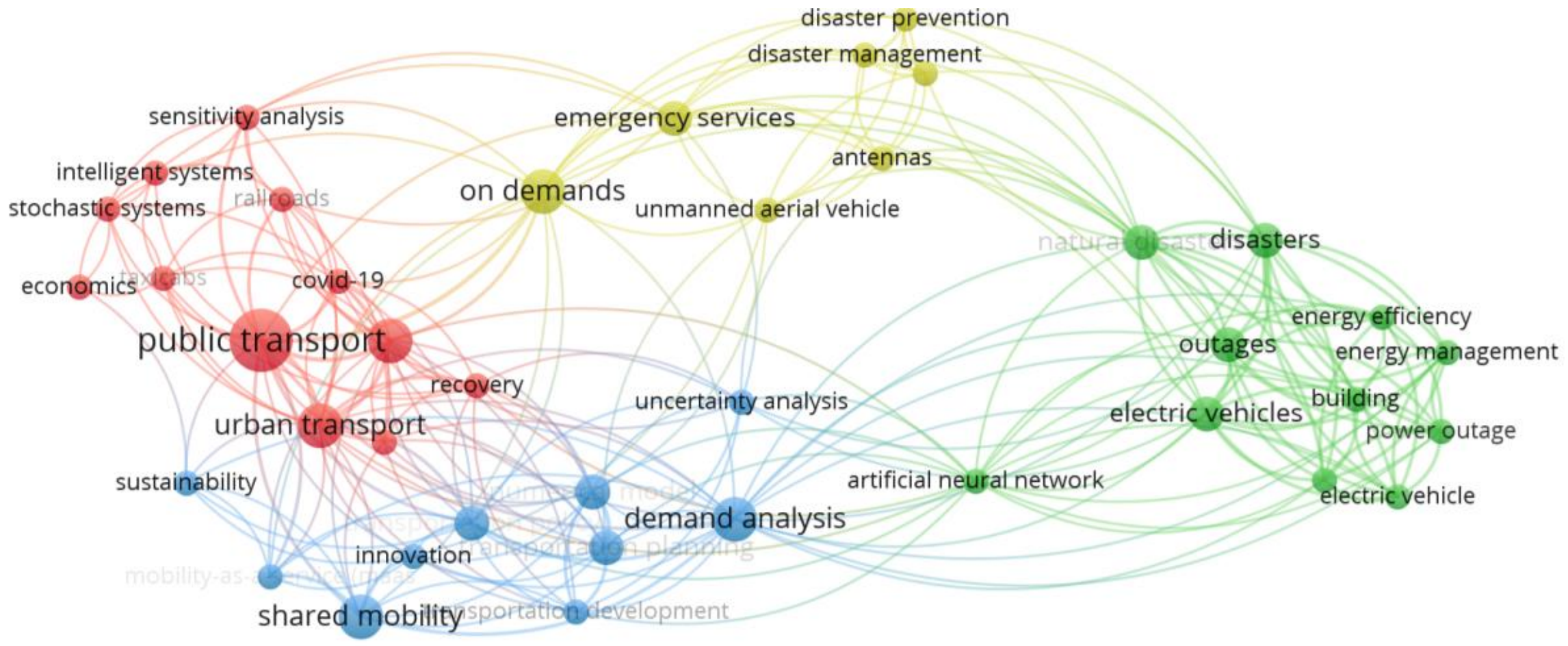

**Figure 9.** Co-occurance network of literature in Theme 1 and 3.

Figure 10 shows the co-occurrence literature network using the keywords from the papers in Theme 2 and Theme 3. Jointly considering Figure 10 with Figure 7 (b) and (c), we see three main groups of concepts: electrification/power outage support (in green), disaster management (in red), and pandemics/COVID-19 (in blue). Furthermore, it is interesting to see that the first and third groups are both associated with the second group, but the first and third groups themselves are not closely associated. On one hand, electrification/power outage are indeed less related topics; on the other hand, such a lack of connectivity might hint potential research on these two topics. For example, it would be interesting to examine the impact of power outages on people's experiences with and without the pandemic and whether electric MOD services can help mitigate the impact of power outages.

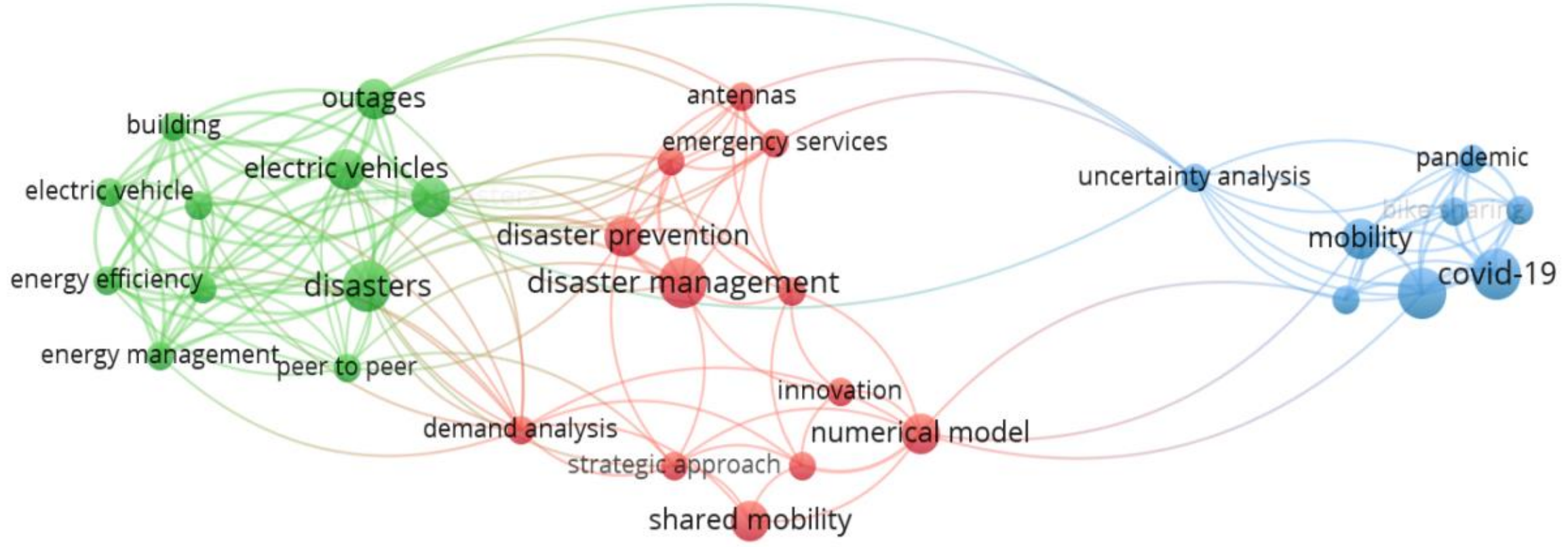

**Figure 10.** Co-occurance network of literature in Theme 2 and 3

Studies in this theme collectively highlight the potential of advanced mobility technologies and novel operational strategies to enhance the resilience of urban infrastructure and communities during disasters. They propose feasible solutions that integrate modern technology with traditional disaster response mechanisms, suggesting a multi-faceted approach to resilience planning that includes technological innovation, social dynamics, and economic considerations. This theme underscores the importance of interdisciplinary research and collaboration across technology developers, urban planners, and disaster response professionals to realize the potential of these innovative mobility solutions. Further empirical and pilot studies are needed to evaluate the practical implementation and effectiveness of these conceptual models in real-world disaster scenarios.

## 4.4 Enabling and Augmenting Technologies (Theme 4)

The literature review within this theme underscores the transformative impact of enabling and automating technologies on MOD-R services, particularly emphasizing their pivotal role in enhancing resilience and functionality in emergency and disaster management scenarios. These studies collectively span a broad spectrum of technological advancements, from communication networks to advanced sensor data processing, decision support systems, and robotics, all of which significantly augment the capabilities of MOD-R services.

A significant focus within this theme is on enhancing ad hoc communication networks through MOD service vehicles, with advancements in mobile wireless ad hoc network routing protocols being particularly notable (Ajrawi and Tran 2024). These studies highlight the importance of robust and efficient communication systems, which are critical for the operational efficacy of MOD services. Moreover, the research delves into cybersecurity aspects, especially defending against large-scale network attacks, underscoring the essential role of cybersecurity in maintaining the safe operation of MOD services (Zhang et al., 2016). The use of UAVs in transmitting live visual data for applications such as disaster management and search and rescue is extensively discussed (Skinnemoen, 2014). This theme also explores the strategic deployment of UAVs to support disaster management efforts, with studies proposing learning-based frameworks to optimize UAV operations for resilient and equitable communication services (Bai et al., 2022). Further research evaluates various routing protocols for UAV-assisted communication networks post-disaster, offering insights into their effective integration into emergency networks to enhance resilience (Choudhary et al. 2024). Denis et al. (2015) present the use of Earth Observation satellites in Europe, highlighting the emergence of shared-ownership or community-owned systems, the civil applications of UAVs, and the development of social media and crowd-based data for humanitarian aid and emergency response.

Innovative applications of virtual reality (VR) combined with drones enhance risk communication associated with catastrophic events, providing immersive approaches to engage the public and decision-makers (Spero et al., 2022). Additionally, the integration of machine learning and AI in vehicle routing, notably in disaster situations, along with virtual reality and digital twin technologies for planning and simulation, is emphasized as a key area that aids in making informed decisions and optimizing routes and modes (Dai et al. 2010; Wolf et al. 2022). Matsuoka et al. (2024) develop an online decision support method for pickup and delivery considering fuel and demand uncertainty, which can be used for both business-as-usual food delivery and disaster rescue.

Research in electrification, such as vehicle-to-grid (V2G) systems, contributes significantly to the sustainability and efficiency of electric MOD services (Sridharan et al., 2023). Moreover, advancements in robotics and mechanical technologies, such as the use of drones for search-and-rescue operations and disaster emergency management, underscore the potential of robotics in enhancing the operational capabilities of MOD services in critical situations (Zhou et al. 2020; Zou et al. 2023).

To better understand the relationship of the papers in Theme 1 and Theme 4, Figure 11 shows the co-occurrence literature network. Jointly considering Figure 11 with Figure 7 (a) and (d), we do not see strong group effects, indicating that papers on R-MOD utilize a wide range of technologies, and a wide range of technologies can be used to improve different aspects of R-MOD. However, it is still interesting to see that communication and computing technologies (in red) are less connected with multimodal services (in blue and green), indicating that the technologies have indeed been used for more for UAVs than other modes of travel services.

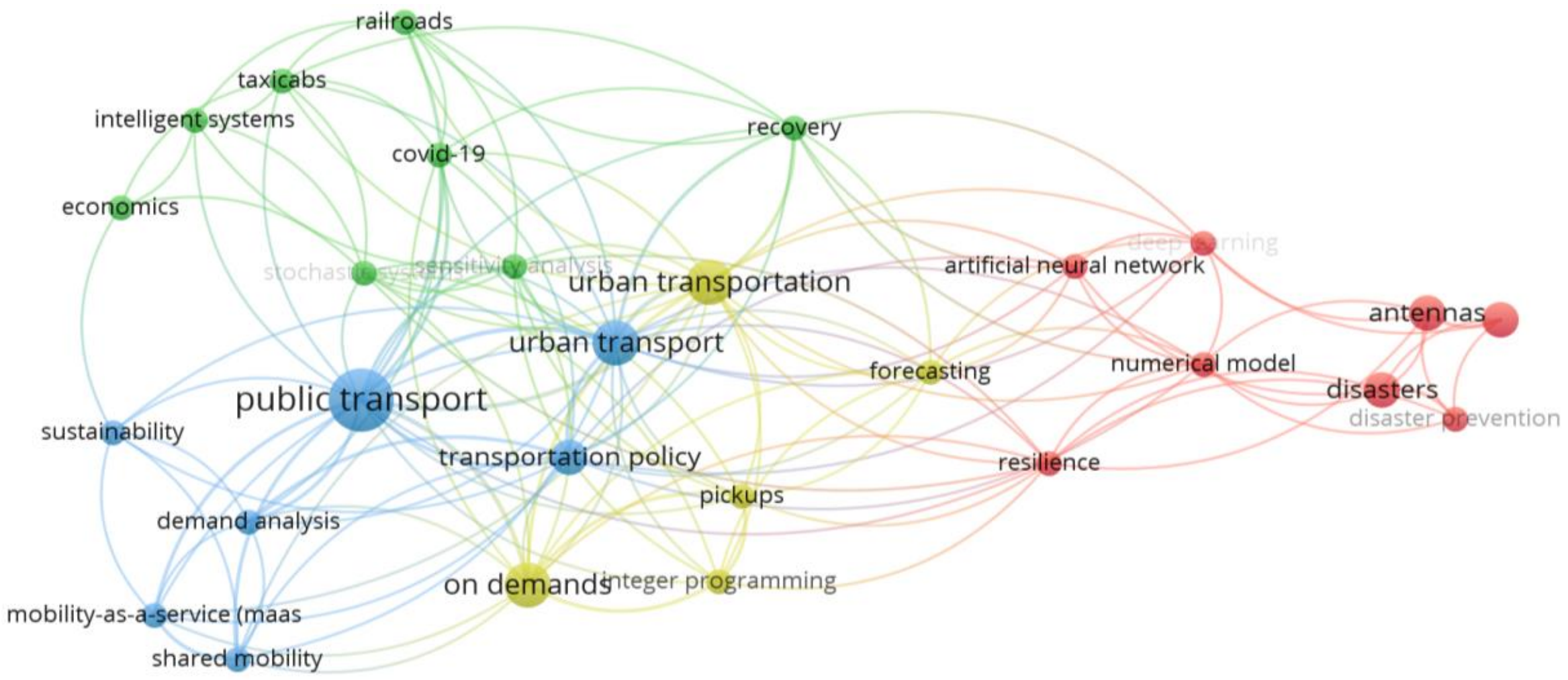

**Figure 11.** Co-occurance network of literature in Theme 1 and 4

Figure 12 shows the co-occurrence literature network for the papers in Theme 2 and Theme 4. Jointly examine Figure 12 with Figure 7 (b) and (d), we do not see strong group effects. The lack of co-occurrence between Theme 2 and 4, especially when comparing Figure 11, indicates the lack of empirical evidence and data support for testing the technologies, techniques, and algorithms for MOD-R services.

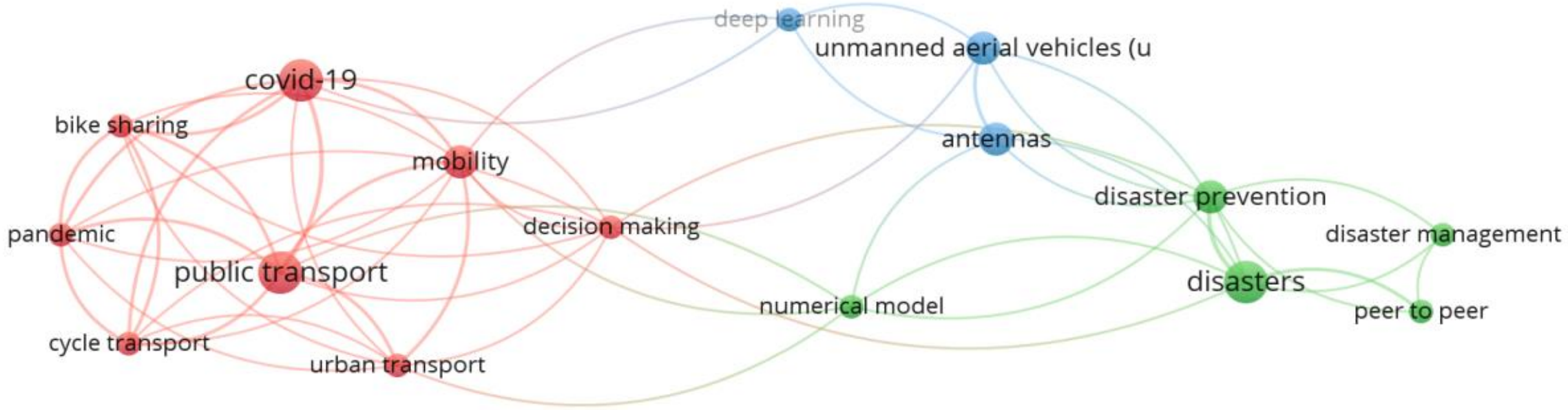

**Figure 12.** Co-occurance network of literature in Theme 2 and 4

To gain an understanding of the relationship of the papers in Theme 3 and Theme 4, Figure 13 shows the co-occurrence literature network of their keywords. Jointly considering Figure 13 with Figure 7 (c) and (d), we see two main groups, connected through the keywords related to "disasters." This aligns with the findings from the trend analysis in Section 3, showing that the pandemics significantly triggered research in MOD-R. The left group (in red) focuses on the use of electric MOD for energy resilience, while the group on the right (in green) focuses on various technologies (e.g., antenna) and techniques (numerical models) in different disaster and demand scenarios.

These technological advancements not only enhance the functionality of MOD-R services but also necessitate a comprehensive framework for integrating these technologies into cohesive MOD systems. Future research should focus on exploring the synergies between these technologies and empirically evaluating their applications in real-world scenarios. Moreover, understanding the interplay between these technologies and human intelligence can lead to more holistic and effective urban resilience strategies.

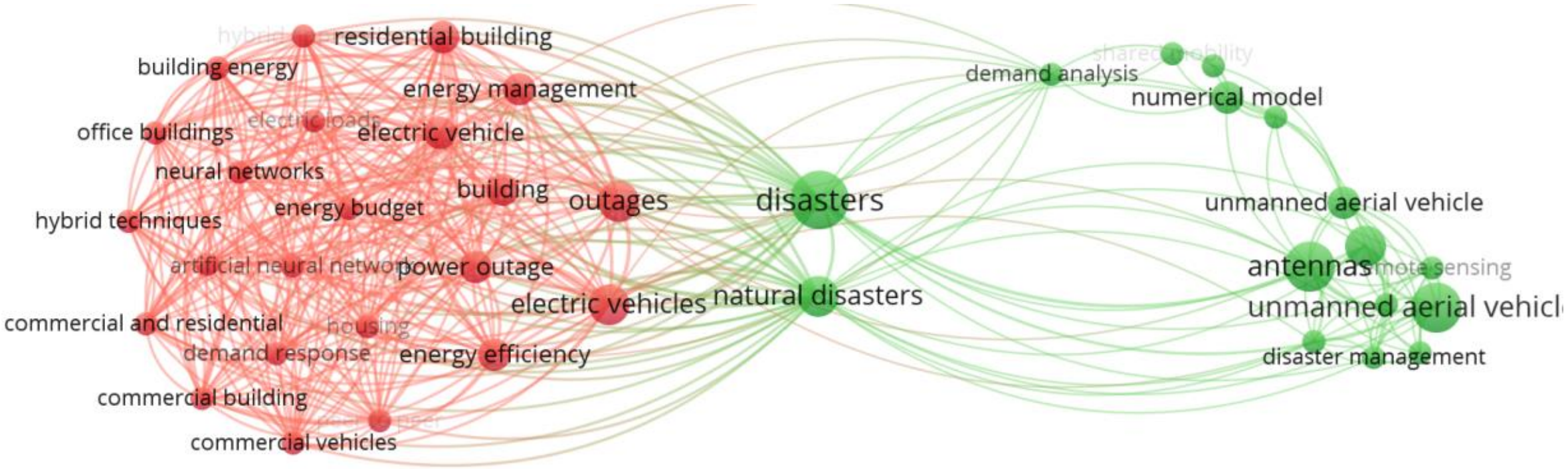

**Figure 13.** Co-occurance network of literature in Theme 3 and 4

## 4.5 Summary and Research Gap

The review across the four themes highlights the importance and the potential benefits of MOD-R services. Table 2 selects some key representative papers to show the wide range of application scenarios, study methodologies, and integration with the rest of the urban mobility systems. The review also highlights the potential of integrating advanced technologies, novel operational strategies, and human-machine collaboration to enhance the adaptability and effectiveness of transportation systems in times of crisis.

Despite advancements, there remains a crucial need for frameworks that integrate various technologies—such as UAVs, ride-sourcing, and mobile ad-hoc networks—with traditional transportation systems. Effective integration would facilitate seamless data sharing and decision-making across platforms, enhancing responsiveness in crisis scenarios. Such frameworks could leverage insights from Xiao et al. (2017), who demonstrated the optimization of resource allocation in disaster response, to ensure efficient and coordinated MOD-R operations. Current literature lacks depth in exploring the psychological resilience of first responders and decision-makers during crises. Detailed studies are needed to understand how stress impacts operational efficiency and decision-making within MOD services. Research in this area could build upon the modeling methods suggested by Yu and Hyland (2020), which consider individual decisions under pressure, to develop systems that support effective decision-making under stress. There is a significant gap in the empirical testing of theoretical models proposed for MOD services. Future research should conduct field tests and simulations to validate these models' effectiveness in real-world crises. Moreover, developing standardized resilience metrics, as discussed by Poulin and Kane (2021) would facilitate the broader implementation and assessment of MOD-R services. The deployment of innovative MOD technologies requires supportive policy frameworks that facilitate rapid technological adaptation during emergencies. Research should focus on governance models that enable such flexibility while addressing the regulatory, ethical, and social implications, as highlighted by Kawasaki and Rhyner (2018).

There is an evident lack of uniform, comprehensive frameworks for evaluating the performance of MOD-R services across various contexts. Future research should aim to develop standardized, multidimensional assessment tools that evaluate these services in terms of efficiency, cost, adaptability, ecological impact, and user satisfaction. Such frameworks are crucial for benchmarking and optimizing MOD-R services, ensuring they meet diverse urban needs while enhancing overall resilience. Investigating how MOD services interact synergistically with both urban infrastructure and the natural environment remains underexplored. Future research should consider the long-term sustainability and resilience impacts of MOD services on urban and ecological systems. This approach aligns with the comprehensive evaluation framework for urban resilience proposed by Datola (2023), which emphasizes the integration of economic, social, environmental, natural, and governance dimensions. Future research may focus on developing frameworks to maximize synergy between different information sources, transportation models, technologies, algorithms, and human-machine interactions, enhancing infrastructure and community resilience. Such frameworks will integrate real-time data from Earth observation satellites and UAVs, as emphasized by Denis et al. (2015), with advanced computational models like those proposed by Xiao et al. (2017) to optimize resource allocation in disaster responses. Additionally, the research will explore the development of a digital twin for urban areas to support multi-agency incident management, drawing on the prototype discussed by Wolf et al. (2022). This digital twin will serve as a dynamic simulation model that assists in visualizing the impact of disaster scenarios and evaluating different response strategies. By addressing these research gaps and exploring advanced integrative approaches, the proposed research aims to significantly advance the field of urban resilience, offering innovative solutions that can be adapted to cities worldwide.

**Table 2.** Summary of the selected representative papers.

| Paper | Title | Disruptive Event | MOD service purpose | Augmenting Technologies | Study Methods |
|---|---|---|---|---|---|
| Ahmed et al. (2020) | Modeling social network influence on hurricane evacuation decision consistency and sharing capacity | Multitype (e.g., hurricanes) | Evacuation | Social network (advanced communication) | Online Survey , Personal Network Research Design (PNRD) framework, Tobi regression. |
| Amirioun et al. (2023) | Resilience-Oriented Scheduling of Shared Autonomous Electric Vehicles: A Cooperation Framework for Electrical Distribution Networks and Transportation Sector | Predictable extreme events such as scheduled network-wide maintainance of power system | Grid power stabilization | Electrification (Vehicle-to-Grid), automation | Linear programming |
| Borowski et al. (2023) | Does ridesourcing respond to unplanned rail disruptions? A natural experiment analysis of mobility resilience and disparity | Unplanned rail disruptions | Substituting transit | Advanced communication (for timing responding to demand surge) | Natural experiment analysis |
| Chokotho et al. (2017) | First responders and prehospital care for road traffic injuries in Malawi | Traffic collisions | On-site (pre-hospital) care | Advanced communication | Focus group |
| Jeong et al. (2020) | The humanitarian flying warehouse | Military conflicts | Essential supply delivery & emergency logistics | Hybrid airbone delivery | Mixed integer programming |
| Kirubakaran and Hosek (2023) | Optimizing Tethered UAV Deployment for On-Demand Connectivity in Disaster Scenarios | Multitype | Ad Hoc communication | Advanced communication (between human operator and UAVs and between robotic cars and UAVs), Automation (Flight control) | Genetic algorithm (on Matlab) |
| Moug et al. (2023) | A shared-mobility-based framework for evacuation planning and operations under forecast uncertainty | Multitype | Evacuation | Advanced communication (for coordinating among vehicles) | Two-stage mixed-integer programming |
| Patil et al. (2022) | Smart UAV Framework for Multi-Assistance | Multitype (flood, pandemic, etc.) | Delivering goods, medical support, rescue, monitoring and detection, evacuation, increasing security, etc. | Advanced communication, Automation (Flight control) | -- |
| Qin and Karimi (2023) | Evolvement patterns of usage in a medium-sized bike-sharing system during the COVID-19 pandemic | Pandemics | Mode substitution | -- | Empirical data analysis (natural experiment) |
| Sridharan et al. (2023) | A hybrid approach based energy management for building resilience against power outage by shared parking station for EVs | Power outage | Power support | Electrification (Vehicle-to-Grid) | EV model, EV-Building model, power flow model, demand response (DR) program, Chicken search optimization algorithm, spike neural network learning algorithm |
| Teixeira et al. (2022) | The strengths and weaknesses of bike sharing as an alternative mode during disruptive public health crisis: A qualitative analysis on the users' motivations during COVID-19 | Pandemics | Transit system support | Communication (for bikesharing system management) | Semi-structured interview |
| Wang et al. (2021) | Modeling and analysis of optimal strategies for leveraging ride-sourcing services in hurricane evacuation | Hurricane | Evacuation | Communication (for ride-sourcing service management) | Two-sided market model that enables subsidy consideration |
| Yu et al. (2023) | Improving Infrastructure and Community Resilience with Shared Autonomous Electric Vehicles (SAEV-R) | Multitype (Earthquake, Pandemics, Hurricane, etc.) | Transport repair and healthcare crew, grid stabilization, essential supply delivery, evacuation, rescue, etc. | Electrification (for Vehicle-to-Grid), automation, advanced communication | System dynamics, state-space compartment model |

# 5 MOD-R Involved Synergistic Humanitarian Logistics

As suggested from Section 4.5, the successful implementation of MOD-R hinges on the synergistic integration of advanced technologies, effective human-machine teaming, and seamless interaction with existing logistics and transport systems. This approach ensures that MOD-R services not only respond efficiently to emergencies but also contribute to the sustainable and resilient growth of urban environments. This section outlines a strategic framework for designing, developing, and operating MOD-R services that capitalize on these synergies, and we refer to these research areas collectively as the lack of synergistic MOD-R design, development, and operation of MOD-R services. Next, we provide forward-looking discussions in three main areas. The proposed synergistic design, development, and operation is illustrated in Figure 14, where bi-level (embedded) feedback loops represent the long-term vision and planning, and the (near) real-time disaster response, respectively.

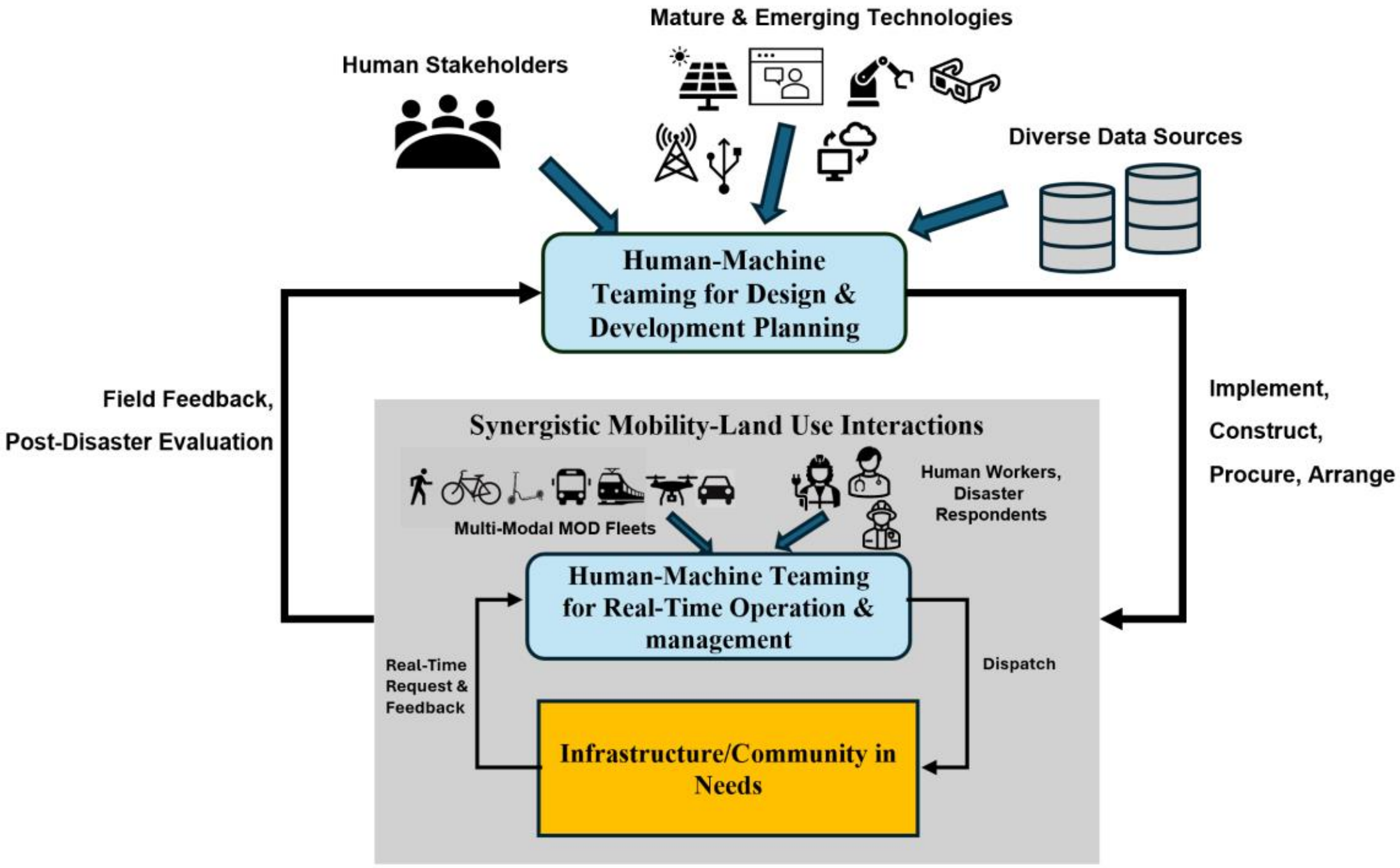

**Figure 14.** Two-Layer (or Embedded) Feeback loops for synergistic design, development, and management of humanitarian MOD-R services.

## 5.1 Integrating Technologies, Modes, and Information Sources

In the rapidly evolving landscape of urban mobility, the integration of mature and advanced technologies and algorithms (Al Ajrawi and Tran, 2024; Liu et al., 2022; Tingstad Jacobsen et al., 2023), different modes of transport (Borowski et al., 2023; Sanchez et al., 2022; Ezaki et al., 2024; Yu et al., 2023), diverse data sources (Barbosa et al., 2018; Lim et al., 2021), holds the key to enhancing multi-modal MOD services. This subsection explores promising research directions that leverage technological synergies and information integration to develop robust, adaptable, and efficient MOD-R services capable of responding to urban disruptions effectively.

Future research can build on the work of Henry et al. (2022), who demonstrated the efficacy of stochastic modeling in optimizing the location of park-and-ride facilities to improve network resilience. Extending this approach, researchers should explore the integration of real-time mobility data and user behavior analytics to dynamically adjust location strategies in response to ongoing urban events or disruptions. Because of the multi-purpose nature of MOD-R, tailored routing algorithms that have flexibilities for pre-scheduled and real-time adjustment in priorities are needed (Worasan et al., 2024). The approach by Wang et al. (2020) utilizing diffusion graph convolutional networks to predict and evaluate transportation resilience under extreme conditions shows significant promise. Future studies should focus on expanding these models to include more diverse data sources, such as social media and crowd-sourced input, to enhance the predictive accuracy and real-time responsiveness of MOD systems. Inspired by the study from Auad et al. (2021) on the resilience of multimodal transit systems during pandemics, further research is needed to design flexible MOD systems that integrate various transport modes. These systems should be capable of adjusting service patterns instantly based on real-time health data and changing public health directives to maintain operational continuity and public safety. The convergence of IoT and AI technologies, as discussed by Liyanage et al. (2019), offers substantial opportunities to advance MOD-R. Future research should focus on developing AI algorithms that can analyze IoT-generated data to make autonomous decisions about vehicle dispatch, route optimization, and service adjustments during disruptions. Reflecting on the insights from Ma et al. (2018) regarding user behavior's impact on service scalability and resilience, further research should explore mechanisms for better governance models that include active user participation. This could involve developing platforms that allow real-time feedback from users, which can be integrated into service planning and operational adjustments. Incorporating advanced telecommunications and real-time data transfer technologies can significantly enhance the coordination and efficiency of MOD services. Research should explore the development of resilient communication networks that remain operational during severe disruptions, ensuring continuous connectivity between vehicles, control centers, and users.

## 5.2 Human-Machine Teaming

The integration of human expertise with advanced machine capabilities is pivotal in enhancing MOD-R services. This synergy is essential not only for operational efficiency but also for ensuring the ethical deployment and societal acceptance of these technologies.

Studies have underscored the importance of incorporating human roles into the deployment of MOD-R services effectively. Chokotho et al. (2017) and Kim and Davidson (2015) highlight the crucial involvement of diverse human participants such as community leaders, police officers, commercial drivers, and social volunteers in emergency responses. These roles ensure regulatory compliance and the effective use of unmanned systems in disaster management. With the advent of autonomous technologies like vehicles and rescue robots, the collaboration between humans and machines becomes increasingly significant. Kirubakaran and Hosek (2023) advocate for the remote human operation of Tethered UAVs during disasters to enhance reliability, while Patil et al. (2022) note the efficiency of AI drones in scanning large areas rapidly, which can prevent human responders from risking their lives. The integration of AI and cloud computing has led to more sophisticated human-machine interfaces, necessitating human oversight to manage and secure these communications, especially in emergency scenarios (El Defrawy and Tsudik, 2008). Moreover, as Wong et al. (2021) discuss, understanding human factors in resource sharing during crises is crucial for designing technology platforms that are not only efficient but also sensitive to human behaviors and social dynamics.

The evolving landscape of MOD-R services necessitates specific human roles and skills development at various stages of disaster management. For example,

- Disaster Response Planners: Professionals like those mentioned by Chang et al. (2014) require robust data analysis and system design skills to strategize infrastructure resilience against natural disasters.
- Disaster Response Operations: As indicated by Chokotho et al. (2017) and Kim and Davidson (2015), rapid decision-making and crisis management expertise are critical for emergency response operators who utilize real-time data and analytics provided by MOD technologies.

- Social and Delivery Workers: The pandemic has shown the benefits of integrating robots for doorstep delivery to minimize human contact, as highlighted by the integration of delivery robots during social distancing measures.
- Infrastructure and Equipment Repair Crew: Highlighted by Yu et al. (2023), the combination of human expertise and robotic assistance is essential for the efficient repair of infrastructure and equipment post-disaster.
- Public Outreach and Communication Staff: Effective communication strategies that consider human behaviors are vital for increasing the acceptance of MOD services, as discussed by Wong et al. (2021).

Lastly, the introduction of generative AI and multi-modal large language models might revolutionize how human-machine interfaces are designed and utilized in MOD-R services (Barreto et al., 2023). These advancements allow for more autonomous decision-making processes where human roles shift towards managing AI operations, interpreting AI-driven insights, and ensuring ethical alignment. As MOD-R services continue to evolve, future research should focus on developing intuitive interfaces that enhance human control over these systems, ensuring that they remain user-friendly and aligned with human operational needs. Additionally, addressing ethical and equity considerations is crucial to ensure equitable access and manage privacy and security concerns effectively. This collaboration between human intelligence and machine capabilities will be key to maximizing the potential of MOD services in improving urban resilience and effectively managing emergencies and disasters. Future studies should continue to explore these evolving roles, focusing on training and skill development, ethical considerations, and the creation of intuitive interfaces for improved human-machine collaboration, offering new avenues for enhancing urban resilience and effective disaster management.

## 5.3 MOD-R Development in, with, and for Urban Environment

The detailed business models for enabling MOD-R often directly determines the success of MOD-R services. The trade-offs between contract-based and on-call sourcing (Kantari et al., 2021) and the pricing mechanism of the procurement of last-mile delivery capacities (Gao et al., 2024) are two examples. The specific operation models will have different system efficiency and equity impacts. Logistics is an important but only part of a sustainable humanitarian supply chain, so the integration of MOD-R into a holistic humanitarian supply chain is also important. Kunz and Gold (2017) introduce a framework that suggests that the design of humanitarian supply chain needs to be aligned not only to relief organizations' enablers, but also to the population's long-term requirements as well as any socio-economic and governmental contingency factors. Liu et al. (2023) introduce a resilience assessment framework that comprehensively analyzes the coupling mechanism, structural and functional characteristics of multimodal public transit networks. Due to the complexity of such supply chains and their interactions of the urban environment, it is also important to develop comprehensive and current modeling methods that consider various uncertainties and factors to compare different alternatives before implementation (Azadeh et al., 2014).

The strategic integration of MOD-R services within urban planning processes is critical. By leveraging the unique features of the natural and built environment, these services can significantly enhance urban resilience. Research and development in this area must focus on providing sufficient infrastructure that enables multi-purpose MOD-R services. For example, when MOD-R services are expected to provide energy supply during power outage, the associated vehicle-to-grid (V2G) technologies need to be sufficiently deployed (Yu et al. 2023; Clement-Nyns et al. 2011). For another example, when utilizing on-demand drone for emergency delivery logistics, appropriate and safety docking stations are needed (Pachayappan and Sundarakani, 2023). The efforts on implementing MOD-R systems should also harmonize with and capitalize on urban landscapes. Urban areas exhibit diverse characteristics that must be considered when embedding MOD-R services into long-term development plans. Indeed, research by Chang et al. (2014) highlights the importance of incorporating resilience strategies that address specific environmental and infrastructural vulnerabilities of urban areas. Future studies should also develop frameworks that integrate MOD-R planning with urban growth models, considering factors such as population density, land use, and climate risk. Such integration ensures that MOD-R services are optimally designed and located to serve both current and future urban dynamics.

The unique physical and socio-economic characteristics of urban areas provide opportunities to customize MOD-R services effectively. For instance, Wang et al. (2021) emphasize the importance of using technology to plan and execute evacuations, suggesting that urban form and layout should guide the deployment strategies of MOD-R services. Dense urban centers may require high-capacity, rapid-response MOD systems, whereas suburban or rural areas might benefit from solutions focused on broader coverage and versatility. Integration with existing infrastructure, such as utilizing public transportation networks for rapid deployment during emergencies, can enhance the efficiency and reach of these services. The dynamic nature of urban environments, especially under stress from natural disasters or human-induced disruptions, calls for adaptive MOD-R technologies. Infrastructure that supports quick adaptability or multipurpose use in emergencies is crucial. For example, Yu et al. (2023) suggest the use of MOD-R services for efficient mobilization of repair crews and equipment to damaged sites, which implies a need for infrastructure that supports such flexibility. Additionally, technologies that adjust operational strategies in real-time based on sensor data from the built environment can dramatically improve the responsiveness of MOD-R services.

# 6 Conclusion

This review systematically explores the existing research efforts related to MOD-R services. The results suggest the promising role of MOD-services in future humanitarian logistics. The close connection with urban resilience strategies underscores a multi-dimensional landscape where technological innovation converges with human expertise and systemic planning. The investigation into resilient MOD services, empirical impact evaluations, novel applications, and enabling technologies elucidates the pivotal role that MOD services can play in enhancing the adaptability and robustness of urban systems. Key findings reveal that MOD services, enhanced by electrification, automation, advanced communication and computing technologies, exhibit significant potential in diverse roles—from anomaly detection to essential supply delivery and infrastructure resilience. These services adapt dynamically to urban demands, particularly during disruptions, thus reinforcing the urban fabric against various crises. The integration of these augmenting technologies into MOD-R not only bolsters the operational capabilities of these services but also catalyzes broader technological advancements in urban environments.

A critical insight from this review is the indispensable synergy between human capabilities and autonomous systems. The effective deployment of MOD services relies profoundly on the integration of human roles—ranging from emergency responders to social workers and communication specialists. These roles ensure the judicious use of technology, emphasizing human judgment, empathy, and strategic decision-making alongside automated efficiency. This human-machine synergy is crucial for fostering a resilient urban ecosystem. Looking forward, the emergence of generative AI and multi-modal large language models presents novel opportunities for enhancing human-machine collaboration within MOD-R services. These technologies promise to extend the capabilities of MOD services beyond traditional operational roles, facilitating more autonomous and sophisticated decision-making processes. As the MOD-R landscape evolves, it will necessitate professionals to develop new competencies in AI management, ethical oversight, and comprehensive urban planning to ensure the responsible and effective use of emerging technologies.

In conclusion, the integration of MOD services within urban resilience frameworks highlights a forward-thinking approach to urban planning. This paper contributes to the academic and practical understanding of how synergistic interactions between technology and human expertise can be harnessed to enhance urban resilience effectively. Future research should continue to explore this integrative approach, aiming to optimize the balance between technological advancements and human insights to manage urban emergencies and enhance disaster preparedness. This exploration is not only crucial for advancing theoretical frameworks but also for informing policy and practical applications in urban resilience strategies.

# Data Availability Statement

We (the authors) have made the data and materials that support the results or analyses presented in their paper freely available upon request.